\documentclass[conference]{IEEEtran}
\usepackage{epsfig,cite,bm}
\def\argmin{\mathop{\mbox{arg\,min}}}
\IEEEoverridecommandlockouts

\begin{document}
\title{Reduced-Feedback Opportunistic Scheduling and Beamforming with GMD for MIMO-OFDMA}
\author{Man-On Pun, Kyeong Jin Kim, Ronald Iltis and H. Vincent Poor
\thanks{M.O. Pun is now with Mitsubishi Electric Research Laboratories (MERL), Cambridge, MA 02139. This work was done when he was at Princeton University.}
\thanks{K.J. Kim is with Nokia Inc, 6000 Connection Drive, Irving TX 75039.}
\thanks{R. Iltis is with the Department of Electrical and Computer Engineering, University of California, Santa Barbara, CA 93106.}
\thanks{H.V. Poor is with the Department of Electrical Engineering, Princeton University, Princeton, NJ 08544.}
\thanks{This research was supported in part by the Croucher Foundation under a post-doctoral fellowship, and in part by the U. S. National Science Foundation under Grant No. ANI-03-38807.}}
\maketitle
\thispagestyle{plain}

% Proc. Asilomar Conference on Signals, Systems, and Computers, Pacific Grove, CA, Nov. 2008

\begin{abstract}
Opportunistic scheduling and beamforming schemes have been proposed previously by the authors for reduced-feedback MIMO-OFDMA downlink systems where the MIMO channel of each subcarrier is decomposed into layered spatial subchannels. It has been demonstrated that significant feedback reduction can be achieved by returning information about only one beamforming matrix (BFM) for all subcarriers from each MT, compared to one BFM for each subcarrier in the conventional schemes. However, since the previously proposed channel decomposition was derived based on singular value decomposition, the resulting system performance is impaired by the subchannels associated with the smallest singular values. To circumvent this obstacle, this work proposes improved opportunistic scheduling and beamforming schemes based on geometric mean decomposition-based channel decomposition. In addition to the inherent advantage in reduced feedback, the proposed schemes can achieve improved system performance by decomposing the MIMO channels into spatial subchannels with more evenly distributed channel gains. Numerical results confirm the effectiveness of the proposed opportunistic scheduling and beamforming schemes.
\end{abstract}

\section{Introduction}\label{sec:intro}
Orthogonal frequency-division multiple-access (OFDMA) has recently emerged as one of the most promising multiplexing techniques for the next-generation broadband wireless communication systems. In an OFDMA downlink system, the base station (BS) simultaneously transmits data to several active mobile terminals (MTs) by modulating each MT's data on an exclusive set of orthogonal subcarriers. In addition to its robustness to multipath fading and high spectral efficiency, OFDMA is particularly attractive due to its flexibility in dynamically allocating subcarriers to different MTs based on their different quality of service (QoS) requirements and channel conditions \cite{Wong99}. Furthermore, the recent advances in multiple-input multiple-output (MIMO) techniques have inspired considerable research interest in MIMO-OFDMA. However, perfect channel state information (CSI) is generally required at the BS in order to fully harvest the advantages provided by MIMO in the downlink transmission, which incurs a formidable amount of feedback from MTs to the BS. To circumvent this obstacle, opportunistic scheduling and beamforming schemes were proposed for single-carrier (SC) systems by exploiting {\em multiuser diversity} with limited channel feedback \cite{Tse02,Chung03}. However, little work has been done on developing opportunistic scheduling schemes for OFDMA. Unlike single-carrier systems, the amount of feedback overhead required in OFDMA grows at a rate proportional to the number of subcarriers, which makes it challenging to design feedback-limited scheduling schemes for OFDMA.

Assuming approximately identical channel fading over adjacent subcarriers, the pioneering work of \cite{Svedman04} devised an opportunistic scheduling scheme for single-input single-output (SISO)-OFDMA by grouping adjacent subcarriers into exclusive clusters and returning only the average signal-to-noise ratio (SNR) of each cluster. The resulting feedback overhead becomes proportional to the number of clusters, rather than the number of subcarriers. However, such feedback reduction is achieved at the price of performance degradation since an increasing cluster size leads to non-negligible channel variations over subcarriers within each cluster. Recently, \cite{Pun08} proposed to resolve this problem by performing beamforming jointly by the BS and MTs. More specifically, the BS employs the same beamforming matrix (BFM) for {\em all} subcarriers while each MT completes the beamforming task for each subcarrier locally. As a result, for a MIMO-OFDMA system with $Q$ subcarriers, \cite{Pun08} requires feedback of one BFM index and $Q$ SNRs  from each MT. However, singular value decomposition (SVD) is employed in \cite{Pun08} to decompose the MIMO channel of each subcarrier into layered spatial subchannels. As a result, the overall bit error rate (BER) performance of the system is usually dictated by the spatial subchannels associated with the smallest singular values.

In this work, we propose improved opportunistic scheduling and beamforming schemes for MIMO-OFDMA downlink systems using geometric mean decomposition (GMD). Since direct power allocation among {\em layered} spatial subchannels is analytically challenging, we propose a GMD-based approach to decompose the MIMO channels into layered spatial subchannels with more evenly distributed signal gains. Since the same BFM is applicable to all subcarriers, the proposed schemes can substantially reduce the required feedback overhead in a fashion similar to \cite{Pun08}. Analytical and simulation results are presented to demonstrate the effectiveness of the proposed schemes.

\underline{Notation}: Vectors and matrices are denoted by boldface letters.  $\left\|\cdot\right\|$ represents the Euclidean norm of the enclosed vector and $\left|\cdot\right|$ denotes the amplitude of the enclosed complex-valued quantity. ${\bm I}_N$ is the $N\times N$ identity matrix and  $\det\left(\cdot\right)$ is the determinant of the enclosed matrix. We use $E\left\{\cdot\right\}$ and $\left(\cdot\right)^H$ for expectation and Hermitian transposition, respectively. Finally, $\left[{\bm A}\right]_{i,j}$ denotes the $i$-th row and $j$-th column entry of matrix $\bm A$ whereas $\left[{\bm a}\right]_{i}$ the $i$-th entry of vector ${\bm a}$.

\section{Signal Model and Background}\label{sec:smodel}
We consider a MIMO-OFDMA downlink system with $Q$ subcarriers and $K$ active MTs. The BS and each MT are equipped with $M$ and $N$ antennas with $M\leq N$, respectively. In this work, we consider a homogeneous network in which the propagation path between each pair of transmit and receive antennas undergoes  independent and identically distributed (i.i.d.) slow frequency-selective fading with a maximum channel length $L$. We denote by ${\bm h}_k^{m,n}(p)$ the $k$-th MT's channel impulse response between its $n$-th receive antenna and the $m$-th transmit antenna during the $p$-th OFDMA block. Assuming that the cyclic prefix (CP) is sufficiently long and synchronization has been achieved, the signal received by the $k$-th MT over its $q$-th subcarrier during the $p$-th OFDMA block can be written as
\begin{equation}\label{eq:rx}
{\bm y}_{k,q}(p)={\bm G}_{k,q}(p)\cdot{\bm x}_{k,q}(p)+{\bm w}_{k,q}(p),
\end{equation}
where ${\bm x}_{k,q}(p)$ is the pre-coded data vector and ${\bm w}_{k,q}(p)$ is a zero-mean circularly symmetric white Gaussian noise with unity variance. Furthermore, ${\bm G}_{k,q}(p)$ is the corresponding frequency-domain channel matrix computed as \cite{kjkim:qr_svd}
\begin{equation}
{\bm G}_{k,q}(p)=\left[\begin{array}{llll}{\bm e}_q^T{\bm h}_k^{1,1}(p)&{\bm e}_q^T{\bm h}_k^{2,1}(p)&\cdots&{\bm e}_q^T{\bm h}_k^{M,1}(p)\\{\bm e}_q^T{\bm h}_k^{1,2}(p)&{\bm e}_q^T{\bm h}_k^{2,2}(p)&\cdots&{\bm e}_q^T{\bm h}_k^{M,2}(p)\\\vdots&\vdots&\ddots&\vdots\\{\bm e}_q^T{\bm h}_k^{1,N}(p)&{\bm e}_q^T{\bm h}_k^{2,N}(p)&\cdots&{\bm e}_q^T{\bm h}^{M,N}(p)\\\end{array}\right],
\end{equation}
with ${\bm e}_i$ being the vector containing the first $L$ elements of the $i$-th column of a $Q$-point discrete Fourier transform (DFT) matrix ${\bm W}$ with $\left[{\bm W}\right]_{\ell,u}=\exp \left(\frac{-j2\pi \ell u}{Q}\right)$ for $0\leq \ell, u \leq Q-1$. In the sequel, we focus on the $k$-th MT over the $p$-th OFDMA block and omit the MT and block indices, i.e. $k$ and $p$, for presentational clarity.

Before proceeding to the proposed schemes, we will first review two straightforward extensions of \cite{Chung03} that were originally developed for SC-MIMO systems. By regarding each subcarrier of OFDMA as an individual system, the MIMO channel of the $q$-th subcarrier can be decomposed into the following form:
\begin{equation}\label{eq:Gqsvd}
{\bm G}_{q}={\bm U}_q\cdot{\bm \Sigma}_q\cdot{\bm V}_q^H,
\end{equation}
where ${\bm U}_q$ and ${\bm V}_q$ are $N\times N$ and $M\times M$ unitary matrices, respectively. Furthermore, ${\bm \Sigma}_q$ is a diagonal matrix whose first $M$ diagonal elements are the singular values of ${\bm G}_{q}$,  $\left\{\lambda_{q,i}\right\}$ for $i=1,2,\cdots,M$. As shown in \cite{Chung03}, ${\bm V}_q$ can be employed as the BFM for the $q$-th subcarrier whereas ${\bm U}_q$ can be employed for data detection at the chosen MT. However, since ${\bm V}_q$ contains the eigen-vectors of ${\bm G}_{q}$ and is {\em dependent} on the subcarrier index $q$, information about all $Q$ BFMs as well as $Q$ real-valued $\left\{\lambda_{q,i}\right\}$ has to be fed back from each MT to the BS. In the sequel, the opportunistic scheduling and beamforming scheme using (\ref{eq:Gqsvd}) is referred to as the per-subcarrier eigen-beamforming scheme (PS-EB).

Inspired by \cite{Svedman04}, the feedback overhead of PS-EB can be reduced by dividing the $Q$ subcarriers into $G$ exclusive clusters with $Q=G\times U$ and returning information about only one BFM and the corresponding average SNR of each cluster. As a result, the feedback overhead is reduced by a factor of $U$. However, since ${\bm V}_n\neq {\bm V}_m$ for $n\neq m$ due to channel variations over subcarriers, using the same BFM for all subcarriers within one cluster leads to system performance degradation. In the sequel, this scheme is referred to as the per-cluster eigen-beamforming scheme (PC-EB).

\section{Per-subcarrier GMD-base scheme (PC-GMD)}
\subsection{MIMO Channel Decomposition}
Rather than directly decomposing ${\bm G}_{q}$ as (\ref{eq:Gqsvd}), we first rewrite ${\bm G}_q$ into the following form:
\begin{eqnarray}
{\bm G}_{q}&=&{\bm W}_q\cdot\left[\begin{array}{llll}{\bm h}^{1,1}&{\bm h}^{2,1}&\cdots&{\bm h}^{M,1}\\{\bm h}^{1,2}&{\bm h}^{2,2}&\cdots&{\bm h}^{M,2}\\\vdots&\vdots&\ddots&\vdots\\{\bm h}^{1,N}&{\bm h}^{2,N}&\cdots&{\bm h}^{M,N}\\\end{array}\right],\\
&=&{\bm W}_q\cdot{\bm H}, \label{eq:GIeH}
\end{eqnarray}
where ${\bm W}_q=\left[{\bm I}_N\otimes{\bm e}_q^T\right]$ with $\otimes$ denoting the Kronecker product.

Next, ${\bm H}$ is decomposed using GMD \cite{Jiang05} as follows:
\begin{equation}\label{eq:Hsvd}
{\bm H}={\bm B}\cdot{\bm E}\cdot{\bm P}^H,
\end{equation}
where ${\bm B}$ and ${\bm P}$ are $NL\times NL$ and $M\times M$ unitary matrices respectively, while ${\bm E}$ is a real-valued upper triangular matrix with its diagonal elements equal to the geometric mean of the positive singular values of ${\bm H}$.

Substituting (\ref{eq:Hsvd}) into (\ref{eq:GIeH}), we have
\begin{eqnarray}
{\bm G}_q&=& {\bm W}_q\cdot{\bm B}\cdot{\bm E}\cdot{\bm P}^H\\
&=&{\bm Q}_q\cdot{\bm R}_q\cdot{\bm P}^H,\label{eq:GQRV}
\end{eqnarray}
where the last equality is obtained by the following QR decomposition:
\begin{equation}
{\bm W}_q\cdot{\bm B}\cdot{\bm E}={\bm Q}_q\cdot{\bm R}_q
\end{equation}
with ${\bm Q}_q$ being an $M\times M$ unitary matrix and ${\bm R}_q$ an $M\times N$ upper triangular matrix.

Inspection of (\ref{eq:GQRV}) suggests that ${\bm P}$ is independent of the subcarrier index $q$ and thus can be employed as the BFM at the BS for all subcarriers, i.e.
\begin{equation}\label{eq:precode}
{\bm x}_q={\bm P}\cdot{\bm s}_q,
\end{equation}
where ${\bm s}_q$ of length $M$ contains the data symbols with
\begin{equation}
E\left\{\left\|{\bm s}_q\right\|^2\right\}=\rho M,
\end{equation}
and $\rho$ being the average SNR.

Furthermore, (\ref{eq:GQRV}) indicates that the beamforming operation for each subcarrier is characterized by ${\bm Q}_q$ which is known to the MT. Recalling (\ref{eq:rx}), (\ref{eq:GQRV}) and (\ref{eq:precode}), data detection at each MT can be carried out by pre-multiplying ${\bm y}$ with ${\bm Q}_q^H$ :
\begin{eqnarray}
{\bm r}_q&=&{\bm Q}_q^H\cdot{\bm y},\\
&=&{\bm R}_q\cdot{\bm P}^H\cdot {\bm P}\cdot{\bm s}_q+{\bm w}'_q,\\
&=&{\bm R}_q\cdot{\bm s}_q+{\bm w}'_q,\label{eq:rQH2}
\end{eqnarray}
where we have exploited the fact that ${\bm Q}_q$ and ${\bm P}$ are unitary matrices and ${\bm w}'_q={\bm Q}_q^H{\bm w}_q$.

Since ${\bm R}_q$ is an upper triangular matrix also known to the MT, ${\bm s}_q$ can be easily detected from (\ref{eq:rQH2}) using existing data detectors such as the QRD-M receiver \cite{Kim08}. In the sequel, this scheme is referred to as the per-subcarrier GMD-based scheme (PS-GMD).

\subsection{Capacity analysis}
It has been shown in \cite{Jiang05} that the adverse effect of error propagation in decoding (\ref{eq:rQH2}) is negligible for the high SNR region. As a result, (\ref{eq:rQH2}) can be rewritten as $M$ equivalent parallel channels:
\begin{equation}
\left[{\bm r}_q\right]_{m}=\left[{\bm R}_q\right]_{m,m}\left[{\bm s}_q\right]_{m}+\left[{\bm w}'_q\right]_{m},\quad m=1,2,\cdots,M.
\end{equation}
Thus, the supportable data throughput on the $q$-th subcarrier is given by
\begin{equation}\label{eq:Cq}
C_q=\sum_{m=1}^M\log_2\left(1+\rho\left|\left[{\bm R}_q\right]_{m,m}\right|^2\right).
\end{equation}

It is interesting to compare the capacity of PS-GMD and PS-EB. Recalling $\left\{{\lambda}_{q,m}\right\}$ are the singular values of the MIMO channel over the $q$-th subcarrier, the corresponding supportable data throughput provided by PS-EB takes the following form:
\begin{equation}\label{eq:Cq}
T_q=\sum_{m=1}^M\log_2\left(1+\rho\left|{\lambda}_{q,m}\right|^2\right).
\end{equation}

Observing that (\ref{eq:GQRV}) and (\ref{eq:Gqsvd}) stand for two different decomposition expressions of the same matrix ${\bm G}_{q}$, we have
\begin{equation}
{\bm Q}_q\cdot{\bm R}_q\cdot{\bm P}^H={\bm U}_q\cdot{\bm \Sigma}_q\cdot{\bm V}_q^H.
\end{equation}

Upon taking the determinant of both sides and recalling that ${\bm Q}_q$, ${\bm P}$, ${\bm U}_q$ and ${\bm V}_q$ are all unitary matrices, we can easily show that
\begin{equation}
\det\left({\bm R}_q\right)=\det\left({\bm \Sigma}_q\right),
\end{equation}
or equivalently,
\begin{equation}
\prod_{m=1}^M \left|\left[{\bm R}_q\right]_{m,m}\right|=\prod_{m=1}^M \left|\lambda_{q,m}\right|.
\end{equation}

Assuming $\rho\left|\lambda_{q,m}\right|\gg 1$ and $\rho\left|\left[{\bm R}_q\right]_{m,m}\right|\gg 1$, we can obtain the following asymptotic relationship between the two data throughputs as $\rho$ increases to infinity.
\begin{equation}
\frac{{C}_q}{{T}_q}\approx\frac{M\log_2\left(\rho\right)+2\log_2\left(\displaystyle\prod_{m=1}^M \left|\left[{\bm R}_q\right]_{m,m}\right|\right)}{M\log_2\left(\rho\right)+2\log_2\left(\displaystyle\prod_{m=1}^M \left|{\lambda}_{q,m}\right|\right)}=1.
\end{equation}
Thus, the data throughput of PS-GMD approaches that of PS-EB asymptotically as $\rho$ increases.

\section{Per-cluster GMD-based scheme (PC-GMD)}
Further feedback reduction can be achieved by dividing the $Q$ subcarriers into $G$ exclusive clusters composed of $U$ adjacent subcarriers with $Q=G\times U$. Assuming that the channel conditions are approximately identical over the $U$ subcarriers in the same cluster, the average supportable data throughput of each cluster, $\bar{C}_g$, is computed and fed back to the BS with
\begin{equation}
\bar{C}_g=\frac{1}{U}\sum_{q\in {\cal I}_g}{C}_q,
\end{equation}
where ${\cal I}_g=\left\{i_{g,1},i_{g,2},\cdots,i_{g,U}\right\}$ is the subcarrier index set of the $g$-th cluster.

As a result, only information about one BFM and $G$ real-valued supportable throughputs are required to be fed back to the BS. In the sequel, this scheme is referred to as the per-cluster reduced-feedback opportunistic scheduling and beamforming scheme (PC-GMD).

\section{Discussion}
\subsection{BFM feedback}
To derive (\ref{eq:rQH2}), we have assumed that ${\bm P}$ is employed as the BFM in (\ref{eq:precode}). However, in practice, BS and MTs share a common BFM codebook with finite entries of unitary matrices. As a result, the MT can only select a unitary matrix that best matches ${\bm P}$ and feed back the index of the chosen BFM to the BS. Denote by $2^B$ the number of entries in the codebook. A viable criterion to select the best BFM, $\tilde{\bm P}_{d^*}$, from the codebook is given by
\begin{equation}\label{eq:dstar}
d^*=\argmin_{d\in\left[1,2^B\right]}\left\|{\bm P}^H\cdot\tilde{\bm P}_d-{\bm I}_M\right\|^2.
\end{equation}

Clearly, there exists a tradeoff between a large codebook and feedback overhead. On the one hand, it is desirable to possess a large codebook such that the chance of achieving ${\bm P}^H\cdot\tilde{\bm P}_{d^*}={\bm I}_M$ is improved. On the other hand, a large codebook requires a large $B$, which in turn incurs more feedback bits. The impact of the codebook size, $B$, on the system performance will be investigated in the next section through computer simulation.

\subsection{Comparison with \cite{Pun08}}
It is instructive to compare the proposed channel decomposition technique shown in (\ref{eq:GQRV}) with PS-QRD proposed in \cite{Pun08}. Recall from \cite{Pun08} that SVD is first applied on ${\bm H}={\bm U}\cdot{\bm \Sigma}\cdot{\bm V}^H$ followed by decomposition of ${\bm W}_q\cdot{\bm U}\cdot{\bm \Sigma}$ into the following QR form:
\begin{equation}
{\bm W}_q\cdot{\bm U}\cdot{\bm \Sigma}=\tilde{\bm Q}_q\cdot\tilde{\bm R}_q\label{eq:SVD},
\end{equation}
where ${\bm U}$ and $\tilde{\bm Q}_q$ are unitary matrices while ${\bm \Sigma}$ and $\tilde{\bm R}_q$ are diagonal and upper triangular matrix, respectively.

Let $\tilde{\bm A}_q=\tilde{\bm Q}_q^H\cdot{\bm W}_q\cdot{\bm U}$, (\ref{eq:SVD}) can be rewritten as
\begin{equation}
\tilde{\bm R}_q=\tilde{\bm A}_q\cdot{\bm \Sigma}\label{eq:SVD2},
\end{equation}

Similarly, letting ${\bm A}_q={\bm Q}_q^H\cdot{\bm W}_q\cdot{\bm B}$, we can rewrite ${\bm R}_q$ in (\ref{eq:GQRV}) as
\begin{equation}
{\bm R}_q={\bm A}_q\cdot{\bm E}\label{eq:GMD2}.
\end{equation}

Since ${\bm G}_q={\bm Q}_q\cdot{\bm R}_q\cdot{\bm P}^H=\tilde{\bm Q}_q\cdot\tilde{\bm R}_q\cdot{\bm V}^H$, we have
\begin{equation}
\det\left({\bm G}_q\right)=\det\left({\bm R}_q\right)=\det\left(\tilde{\bm R}_q\right),
\end{equation}
and subsequently,
\begin{equation}
\prod_{m=1}^M\left|\left[{\bm R}_q\right]_{m,m}\right|=\prod_{m=1}^M\left|\left[\tilde{\bm R}_q\right]_{m,m}\right|,
\end{equation}
where we have exploited the fact that ${\bm Q}_q$, ${\bm P}$, $\tilde{\bm Q}_q$ and ${\bm V}$ are unitary matrices while ${\bm R}_q$ and  $\tilde{\bm R}_q$ are upper triangular matrices.

Furthermore, it is easy to show that $\tilde{\bm A}_q$ and ${\bm A}_q$ are also upper triangular matrices. Thus, we have
\begin{equation}
\left[{\bm R}_q\right]_{m,m}=\left[{\bm A}_q\right]_{m,m}\left[{\bm E}\right]_{m,m},
\end{equation}
and
\begin{equation}
\left[\tilde{\bm R}_q\right]_{m,m}=\left[\tilde{\bm A}_q\right]_{m,m}\left[{\bm \Sigma}\right]_{m,m},
\end{equation}
for $m=1,2,\cdots,M$.

Since $\tilde{\bm A}_q$ and ${\bm A}_q$ are products of the same matrix ${\bm W}_q$ with two different unitary matrices, it is reasonable to argue that their diagonal elements follow the same distribution and have approximately the same values. However, in contrast to ${\bm \Sigma}$ with different eigenvalues on its diagonal, ${\bm E}$ has {\em identical} diagonal elements. As a result, we argue that it is more likely for $\left[{\bm R}_q\right]_{m,m}$ to be approximately equal for $m=1,2,\cdots,M$.

\subsection{Comparison with PS/PC-EB}
Compared to PS/PC-EB, the total amount of feedback overhead of PS/PC-GMD is reduced by half. More importantly, since information about only one BFM is fed back per cluster in PC-EB, the chosen BFM may become a poor approximation of the true eigen-beamforming matrices for subcarriers in the cluster. As a result, severe system degradation is incurred as the cluster size increases. In contrast, PC-GMD requires feedback about only one BFM for all subcarriers, regardless of the cluster size. Consequently, the performance degradation due to clustering in PC-GMD is marginal.

However, it is fair to say that the substantial feedback reduction of PS/PC-GMD is achieved at the cost of higher computational complexity compared to PS/PC-EB since the QRD-M receivers have to be employed to detect the data symbols. Furthermore, some BER performance degradation may be incurred due to the noisy detection errors.

\section{Simulation Results}
Computer simulations are performed in this section to confirm the performance of the proposed scheduling and beamforming schemes. The simulated OFDMA system has $Q=64$ subcarriers and $M=N=2$ antennas. The channel response of each pair of transmit and receive antennas is generated according to the channel model specified in \cite{80211n}. Furthermore, the soft-QRD-M receiver proposed in \cite{Kim08} is employed for data detection with the QRD-M factor equal to $12$ after noise whitening. Since we have assumed a homogenous network, the scheduling scheme that allocates the subcarrier/cluster to the MT with the highest throughput is fair. Finally, the source binary data is coded by a low-density parity-check code (LDPC) encoder with $R=7/8$ before being mapped to a quadrature phase-shift keying (QPSK) constellation for all schemes under investigation.

\subsection*{Case 1: BER performance comparison}
\begin{figure}[htp]
\begin{center}
\includegraphics[scale=0.48]{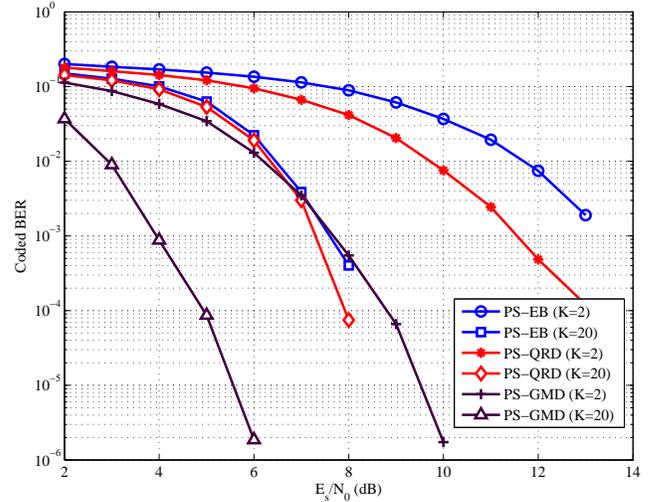}
\caption{BER performance as a function of SNR with $K=10$.}\label{fig:bermfig}
\end{center}
\end{figure}

In the first example, we compare the BER performance of PS-GMD, PS-QRD and PS-EB. Fig. \ref{fig:bermfig} shows their coded BER performance as a function of $E_b/N_0$.

%%%%%%%%%%%%%%%%%%%%%%%%%%%%%%%%%%%%%%%%%%%%%%%%%%%%%%%%%%%%%%%%%
%%%%%%%%%%%%%%%%%%%%%%%%%%%%%%%%%%%%%%%%%%%%%%%%%%%%%%%%%%%%%%%%%

Inspection of Fig. \ref{fig:bermfig} suggests that PS-QRD and PS-EB have approximately the same BER performance for large $K$. This is because their performance is susceptible to the subchannels associated with the small singular values. In contrast, PS-GMD provides good BER performance by benefiting from the subchannels with more evenly distributed channel gains.

\subsection*{Case 2: Impact of codebook size}
In this example, we investigate the impact of the codebook size on coded BER performance. As $B$ increases, the performance of PS-GMD improves. Interestingly, Fig. \ref{fig:gefffig} suggests that PS-GMD with a smaller codebook outperforms PS-EB and PS-QRD with $B=\infty$:

%%%%%%%%%%%%%%%%%%%%%%%%%%%%%%%%%%%%%%%%%%%%%%%%%%%%%%%%%%%%%%%%%
\begin{figure}[htp]
\begin{center}
\includegraphics[scale=0.48]{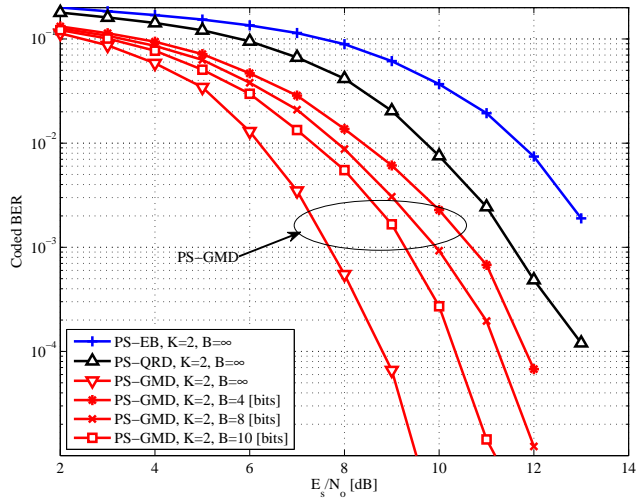}
\caption{Impact of the codebook size, $B$, on coded BER performance}\label{fig:gefffig}
\end{center}
\end{figure}
%%%%%%%%%%%%%%%%%%%%%%%%%%%%%%%%%%%%%%%%%%%%%%%%%%%%%%%%%%%%%%%%%

\begin{figure}[htp]
\begin{center}
\includegraphics[scale=0.48]{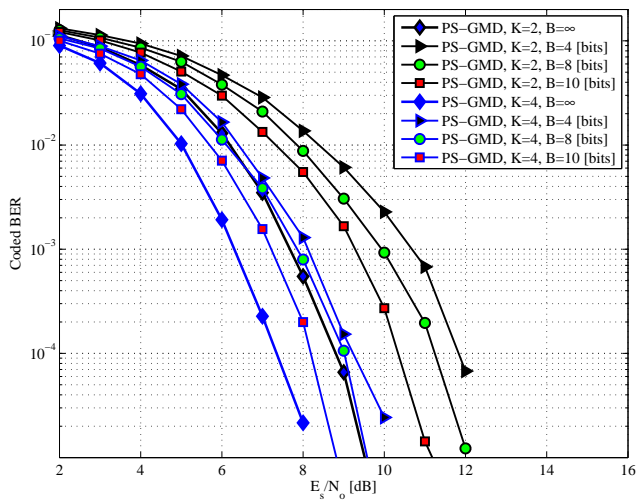}
\caption{Impact of $K$ and $B$ on coded BER performance}\label{fig:fig4}
\end{center}
\end{figure}

Fig. \ref{fig:fig4} shows the coded BER performance of PS-GMD with different $K$ and $B$.
It is evident from Fig. \ref{fig:fig4} that the system degradation due to a smaller codebook
can be alleviated by exploiting more multiuser diversity (i.e. more users) with PS-GMD.
%%%%%%%%%%%%%%%%%%%%%%%%%%%%%%%%%%%%%%%%%%%%%%%%%%%%%%%%%%%%%%%%%
%%%%%%%%%%%%%%%%%%%%%%%%%%%%%%%%%%%%%%%%%%%%%%%%%%%%%%%%%%%%%%%%%

\subsection*{Case 3: Impact of cluster size}
\begin{figure}[htp]
\begin{center}
\includegraphics[scale=0.48]{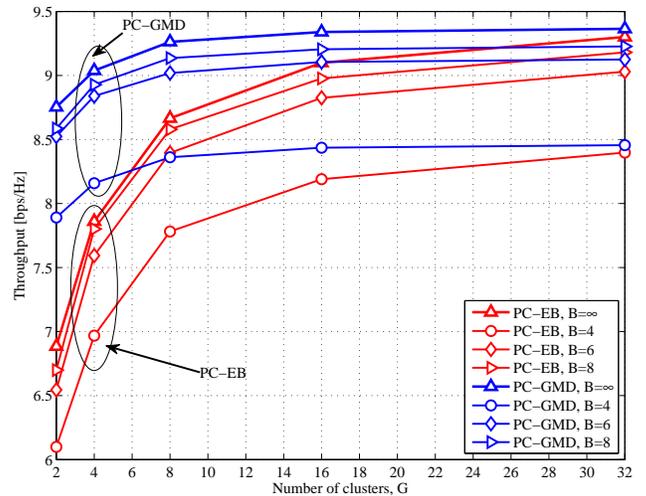}
\caption{Average throughput as a function of the number of clusters, $G$.}\label{fig:fig3}
\end{center}
\end{figure}

In the last example, we compare the throughput performance of PC-GMD and PC-EB in terms of throughput performance with different cluster sizes ($G=2,4,8,16,32$) at an SNR of $10$ dB. As $G$ increases, more accurate channel information is returned to the BS. As a result, the throughput performance increases for both schemes. However, for small $G$, channel variations within clusters become non-negligible. As a result, the BFM chosen by PC-EB to serve the whole cluster becomes a poor approximation of the ideal BFM for each subcarrier in the cluster, which incurs throughput degradation. In contrast, PC-GMD does not entail feedback loss regarding the BFM information, which leads to only marginal performance degradation as observed in Fig. \ref{fig:fig3}, even for small $G$.

%%%%%%%%%%%%%%%%%%%%%%%%%%%%%%%%%%%%%%%%%%%%%%%%%%%%%%%%%%%%%%%%%
%%%%%%%%%%%%%%%%%%%%%%%%%%%%%%%%%%%%%%%%%%%%%%%%%%%%%%%%%%%%%%%%%

\section{Conclusion}
Two opportunistic scheduling and beamforming schemes, namely PS-GMD and PC-GMD, have been proposed for MIMO-OFDMA downlink systems by exploiting a novel beamforming technique where beamforming is performed jointly by both the BS and MTs. The resulting PS-GMD scheme requires feedback of only one BFM and $Q$ supportable throughputs for a MIMO-OFDMA system with $Q$ subcarriers. A further feedback reduction is achieved by PC-GMD through grouping adjacent subcarriers into clusters and returning only cluster information from each MT. Since PC-GMD does not incur feedback loss regarding the BFM information, it only entails marginal performance degradation with respect to PS-GMD. It has been confirmed through asymptotic analysis and computer simulation that the proposed schemes can achieve good BER and throughput performance with substantially reduced feedback.

\bibliographystyle{IEEEtran}
\bibliography{Bib}
\end{document}